\documentclass[pre,twocolumn,showpacs]{revtex4}
\usepackage{graphicx}

\newcommand{\equ}{Eq.}
\newcommand{\eqs}{Eqs.}
\newcommand{\fig}{Fig.}

\newcommand{\sect}{Sec.}
\newcommand{\rem}[1]{}
\newcommand{\const}{\text{const}}
\newcommand{\imag}[1]{\text{Im}(#1)}
\newcommand{\imagb}[1]{\text{Im}[#1]}
\newcommand{\real}[1]{\text{Re}(#1)}
\newcommand{\realb}[1]{\text{Re}[#1]}

\begin{document}

\title{Boundary element method for resonances in dielectric microcavities}    
\author{Jan Wiersig}
\affiliation{Max-Planck-Institut f\"ur Physik komplexer Systeme, D-01187 Dresden, Germany}
\date{\today}
\email{jwiersig@mpipks-dresden.mpg.de}

\begin{abstract} 
A boundary element method based on a Green's function technique is introduced
to compute resonances with intermediate lifetimes in quasi-two-dimensional dielectric cavities.  
It can be applied to single or several optical resonators of arbitrary
shape, including corners, for both TM and TE polarization. 
For cavities with symmetries a symmetry reduction is described. 
The existence of spurious solutions is discussed. 
The efficiency of the method is demonstrated by calculating resonances in two
coupled hexagonal cavities.
\end{abstract}
\maketitle

\section{Introduction}
Dielectric cavities have recently attracted considerable attention due to
the fabrication of microlasers~\cite{ND97,GCNNSFSC98}.
Various shapes have been studied both experimentally and theoretically:  
deformed spheres~\cite{MNCSC95,CCSN00,LW01}, deformed
disks~\cite{ND95,ND97,GCNNSFSC98,CCSN00,SJNS00,GHSSFG00,SOS01,HR02,LLCMKA02,RTSCS02},
squares~\cite{PCC01} and hexagons~\cite{VKLISLA98,BILNSSVWW00}.  
An efficient numerical strategy to compute optical properties of
effectively two-dimensional dielectric cavities with more complex geometries
is the subject of the present paper.  

Maxwell's equations simplify to a two-dimensional (reduced) wave
equation~\cite{Jackson83}   
\begin{equation}\label{eq:wave}
-\nabla^2\psi = n^2({\bf r})k^2\psi \ ,
\end{equation}
with coordinates ${\bf r} = (x,y) = (r\cos{\theta},r\sin{\theta})$, piece-wise
constant index of refraction $n({\bf r})$, (vacuum) wave number $k=\omega/c$,
angular frequency $\omega$ and speed of light in vacuum $c$. 
In the case of TM polarization, the complex-valued wave function $\psi$
represents the 
$z$-component of the electric field vector $E_z({\bf r},t) = \realb{\psi({\bf
r})\exp{(-i\omega t)}}$ with $i^2=-1$, whereas for TE polarization,
$\psi$ represents the $z$-component of the magnetic field vector $H_z$.

The boundary conditions at infinity are determined by the experimental
situation. In a scattering experiment the wave function is composed of an
incoming plane wave with wave vector ${\bf k}$ and an outgoing scattered
wave. The wave function has the asymptotic form (in 2D)
\begin{equation}\label{eq:scatteringbc}
\psi \sim \psi_{\text{in}} + \psi_{\text{out}} = 
\exp{(i{\bf k}{\bf r})} 
+ f(\theta,{\bf k})\frac{\exp{(ikr)}}{\sqrt{r}} \ ,
\end{equation}
where $k=|{\bf k}|$ and $f(\theta,{\bf k})$ is the angle-dependent
differential amplitude for elastic scattering.
In lasers, however, the radiation is generated within the cavity without 
incoming wave,
\begin{equation}\label{eq:outgoingbc}
\psi \sim \psi_{\text{out}} = 
h(\theta,k)\frac{\exp{(ikr)}}{\sqrt{r}} \ .
\end{equation}
This situation can be modelled by a dielectric cavity with complex-valued $n$ leading
to steady-state solutions of the wave equation~(\ref{eq:wave}). Alternatively,
one can use real-valued $n$ leading to states that are exponentially decaying in time. 
The lifetime $\tau$ of these so-called resonant states or short {\it
resonances} is given by the imaginary part of the wave number as 
$\tau=-1/2c\,\imag{k}$ with $\imag{k} < 0$. $\tau$ is related to the quality
factor $Q = \real{\omega}\tau$. 
The resonant states are connected to the peak 
structure in scattering spectra; see~\cite{Landau96} for an introduction. 
Resonant states have been introduced by Gamow~\cite{Gamow28} and
by Kapur and Peirles~\cite{KP38}.   

The wave equation~(\ref{eq:wave}) with the outgoing-wave
condition~(\ref{eq:outgoingbc}) can be solved analytically by means of
separation of 
variables only for special geometries, like the isolated circular cavity (see
e.g. Ref.~\cite{BarberHill90}) and the symmetric annular
cavity~\cite{HR02}. In general, numerical methods are needed.  
Frequently used are wave-matching methods~\cite{ND95}. The wave function
is usually expanded in integer Bessel functions inside the cavity
and in Hankel functions of first kind outside, so that the outgoing-wave
condition~(\ref{eq:outgoingbc}) is fulfilled automatically. 
The Rayleigh hypothesis asserts that such an expansion is always
possible. However, it can fail for geometries which are not sufficiently small
deformations of a circular cavity~\cite{BergFokkema79}. 
It should be mentioned that for a different kind of boundary 
conditions at infinity, the wave-matching method can work well for special 
strongly noncircular geometries, e.g. rectangular integrated 
microresonators~\cite{Lohmeyer02}.

More flexible are, for example, finite-difference methods; see
e.g.~\cite{CZ92}. These methods involve a discretization of the 
two-dimensional space, which is a heavy numerical task for highly-excited
states. An even more severe 
restriction is that it is impossible to discretize to infinity. One has to
select a cut-off at some arbitrary distance from the cavities and implement
there the outgoing-wave condition~(\ref{eq:outgoingbc}). For these reasons, 
finite-difference methods are not suitable for computing
resonances in dielectric cavities.    

A class of flexible methods with better numerical efficiency are 
boundary element methods (BEMs). The central idea is to replace
two-dimensional differential equations such as \equ~(\ref{eq:wave}) by
one-dimensional boundary integral equations (BIEs) and then to discretize the
boundaries. 
BEMs have been widely applied to geometries with Dirichlet boundary conditions
(wave function vanishes), Neumann boundary conditions (normal derivative
vanishes) and combinations of
them~\cite{Kitahara85,CB91,CZ92,Banerjee94}. Bounded states have been
calculated in the context of quantum chaos; for an introduction see
Refs.~\cite{KS97,Baecker02}.   
For scattering problems consider, for example, Ref.~\cite{BM71}. Resonances
have been computed for scattering at three disks by Gaspard and
Rice~\cite{GR89}.  

The boundary conditions for dielectric cavities, however, are of a different 
kind: the wave function and its (weighted) normal derivative are continuous
across a cavity boundary. An analogous quantum problem in semiconductor
nanostructures has been treated by Knipp and Reinecke~\cite{KR96}. Their BEM
is for bounded and scattering states only.
The aim of the present paper is to extend their approach to resonances in
dielectric cavities for TM and TE polarization, including a discussion of
spurious solutions, treatment of cavities with symmetries and cavities with
corners.   

The paper is organized as follows. The BIEs are derived in the framework of
the Green's function technique in \sect~\ref{sec:bie}. Section~\ref{sec:bem}
describes the BEM. Section~\ref{sec:example} demonstrates the efficiency of
the method with an example of two coupled hexagonal resonators. Finally,
\sect~\ref{sec:summary} contains a summary. 

\section{Boundary integral equations}
\label{sec:bie}
\begin{figure}[ht]
\includegraphics[width=5.0cm,angle=0]{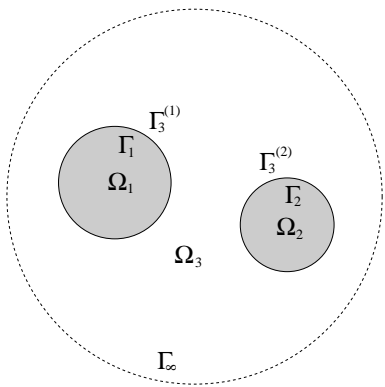}
\caption[]{\footnotesize Geometry and notation for the BIEs. The cavity
with domain $\Omega_1$ is bounded by the curve $\Gamma_1$, the one with
domain $\Omega_2$ is bounded by $\Gamma_2$. The domain $\Omega_3$ is 
``bounded'' by $\Gamma^{(1)}_3 = \Gamma_1$, $\Gamma^{(2)}_3 =
\Gamma_2$ and by a circle $\Gamma_\infty$ at a large distance.}
\label{fig:domains}
\end{figure}

In this section we derive the BIEs for the general case of $J-1$ optical 
cavities in an outer unbounded medium.  
As illustrated in \fig~\ref{fig:domains}, the space is divided into $J$
regions $\Omega_j$, $j=1,2,\ldots,J$, in each of which the index of refraction $n({\bf r})=n_j$
is uniform. Without loss of generality $n_J$ is set to unity, i.e. the
environment is vacuum or air. 
We first concentrate on TM polarization where both the wave function
$\psi$ and its normal derivative are continuous across an interface
separating two different regions. 

To reduce the two-dimensional differential equation~(\ref{eq:wave}) to 
one-dimensional integral equations, we first introduce the Green's function, 
which is defined as solution of 
\begin{equation}\label{eq:green1}
[\nabla^2+n_j^2k^2]G({\bf r},{\bf r}';k) = \delta({\bf r}-{\bf r}') 
\ , 
\end{equation}
where $\delta({\bf r}-{\bf r}')$ is the two-dimensional Dirac
$\delta$-function, ${\bf r}$ and ${\bf r}'$ are arbitrary points within
$\Omega_j$. The outgoing solution for the 
Green's function is   
\begin{equation}\label{eq:green2}
G({\bf r},{\bf r}';k) = -\frac{i}{4}H_0^{(1)}(n_jk|{\bf r}-{\bf r}'|) \ .
\end{equation}
$H_0^{(1)}$ is the zeroth order Hankel function of first
kind~\cite{GradRyzh65}. 

Multiplying the $\psi$-equation~(\ref{eq:wave}) by $G({\bf r},{\bf
r}';k)$ and subtracting the resulting equation from the
$G$-equation~(\ref{eq:green1}) multiplied by $\psi({\bf r})$ gives 
\begin{eqnarray}
\nonumber
\psi({\bf r})\delta({\bf r}-{\bf r}') & = & \psi({\bf r})\nabla^2 
G({\bf r},{\bf r}';k)- G({\bf r},{\bf r}';k)\nabla^2 \psi({\bf r}) \\
\nonumber
& = & \nabla[\psi({\bf r})\nabla 
G({\bf r},{\bf r}';k)- G({\bf r},{\bf r}';k)\nabla \psi({\bf r})] \ .
\end{eqnarray}
Integrating this equation over the region $\Omega_j$ yields on
the l.h.s. $\psi({\bf r}')$ since ${\bf r}'\in \Omega_j$. Applying Green's
theorem, the integral on the r.h.s. can be expressed by a line integral along
the boundary curve $\Gamma_j = \partial\Omega_j$, such that
\begin{equation}\label{eq:bim}
\psi({\bf r}')= \oint_{\Gamma_j}ds[\psi(s)\partial_\nu
G(s,{\bf r}';k)- G(s,{\bf r}';k)\partial_\nu \psi(s)] \ .
\end{equation}
Note that the boundary curve may consist of a number of
disconnected components $\Gamma_j = \Gamma^{(1)}_j \cup \Gamma^{(2)}_j \cup \ldots$ as depicted
in \fig~\ref{fig:domains}. Each component is assumed to be oriented 
counterclockwise, smooth, and not to be a part of $\Omega_j$ itself,
i.e. $\Omega_j$ is an open set.  
$\partial_\nu$ is the normal derivative defined as $\partial_\nu
= {\bf \nu}({\bf r}) \nabla|_{\bf r}$; ${\bf \nu}({\bf r})$ is the outward
normal unit vector to $\Gamma_j$ at point ${\bf r}$;
$s = s({\bf r})$ is the arc length along $\Gamma_j$ at ${\bf r}$. 
The derivative of the Green's function is given by
\begin{equation}\label{eq:normalgreen}
\partial_\nu G({\bf s},{\bf r}';k) = 
\frac{in_jk}{4} \cos{\alpha} \, 
H_1^{(1)}(n_jk|{\bf r}-{\bf r}'|) \ ,
\end{equation}
where $H_1^{(1)}$ is the first order Hankel function of first
kind~\cite{GradRyzh65} and
\begin{equation}
\cos{\alpha} = {\bf \nu}({\bf r}) \frac{{\bf r}-{\bf r}'}{|{\bf r}-{\bf r}'|}
\ .
\end{equation}

The limit ${\bf r}'\to\Gamma_j$ in \equ~(\ref{eq:bim}) is not trivial since
both the Green's function and its normal derivative are singular at ${\bf r}'
= {\bf r}$. However, it can be shown that these singularities are integrable
for smooth boundaries. This is obvious for the second part of the 
integral kernel in \equ~(\ref{eq:bim}) since for small arguments $z=n_jk|{\bf
r}-{\bf r}'|$ 
\begin{equation}\label{eq:H10small}
H^{(1)}_0(z) \sim \frac{2i}{\pi}\ln{z} \ .
\end{equation}
The first part is also integrable. At first glance, this seems to be
surprising because for small arguments 
\begin{equation}\label{eq:H11small}
H^{(1)}_1(z) \sim -\frac{2i}{\pi z} \ .
\end{equation}
However, this singularity is compensated by 
\begin{equation}\label{eq:cosalphakappa}
\cos{\alpha} \sim \frac{1}{2}\kappa |{\bf r}-{\bf r}'| \ ,
\end{equation}
where $\kappa$ is the curvature of the curve $\Gamma_j$ at ${\bf r}(s)$, which is finite for a smooth boundary.
The limit ${\bf r}'\to\Gamma_j$ in \equ~(\ref{eq:bim}) can be performed in the
sense of Cauchy's principal value, see e.g. Ref.~\cite{KS97}, giving 
\begin{equation}\label{eq:cauchybim}
\frac12\psi({\bf r}')= {\cal P}\oint_{\Gamma_j}ds[\psi(s)\partial_\nu
G(s,{\bf r}';k)- G(s,{\bf r}';k)\partial_\nu \psi(s)] \ .
\end{equation}
Comparing the l.h.s of \eqs~(\ref{eq:bim}) and (\ref{eq:cauchybim}) shows
that ${\bf r}' \in \Gamma_j$ gives the ``average'' of the results for ${\bf
r}' \in \Omega_j$ and ${\bf r}' \in \Omega_i$ with $i\neq j$.  

For each region $\Omega_j$ there is an equation as
\equ~(\ref{eq:cauchybim}). Special attention has to be paid to the unbounded
region $\Omega_J$. It is convenient to consider instead a finite region
bounded by a circle $\Gamma_\infty$ with a very large radius $r$ as sketched in
\fig~\ref{fig:domains}. 
We distinguish three cases in the following subsections.

\subsection{Bounded quantum states}
\label{sec:boundedstates}
The case of bounded states in the quantum analogue has been studied by
Knipp and Reinecke~\cite{KR96}. Then, $n_jk$ has to be replaced by
$[2m(E-V_j)]^{1/2}/\hbar$, where $E$ is the energy, $V_j$ with $j=1,\ldots,J$
is a piece-wise constant potential, and $\hbar$ is Planck's constant divided by
$2\pi$. The wave function and its normal
derivative (weighted with the inverse of the effective mass $m$) are 
continuous at domain boundaries. 
If $E < V_J$ then the state is bounded, the wave function and its
gradient vanish exponentially as $r\to\infty$. Moreover, with
$\imag{k} = 0$ the Green's function~(\ref{eq:green2}) vanishes as either ${\bf
r}$ or ${\bf r}'$ goes to infinity. As a result $\Gamma_\infty$ does not
contribute to any of the BIEs. Note that \equ~(\ref{eq:wave}) does not permit  
bounded states since $n_j^2k^2 > 0$. 

Using the same notation as Knipp and Reinecke~\cite{KR96} we reformulate
\equ~(\ref{eq:cauchybim}) as a linear homogeneous BIE 
\begin{equation}\label{eq:hbim}
\oint_{\Gamma_j}ds[B(s',s)\phi(s)+C(s',s)\psi(s)] = 0 \ ,
\end{equation}
with $B(s',s) = -2G(s,s';k)$, $C(s',s) = 2\partial_\nu
G(s,s';k)-\delta(s-s')$, and $\phi(s) = \partial_\nu\psi({\bf r})$.
The entire set of BIEs can be written in a symbolic way as
\begin{equation}\label{eq:symbolic}
\left(\begin{array}{cc}
  B_1 & C_1\\
  B_2 & C_2\\
  \vdots & \vdots\\
  B_J & C_J
 \end{array}
\right)
\left(\begin{array}{c}
  \phi\\
  \psi
 \end{array}
\right) = 
M\left(\begin{array}{c}
  \phi\\
  \psi
 \end{array}
\right) = 0 \ ,
\end{equation}
where $B_j$ and $C_j$ represent the integral operators in region $\Omega_j$. 
The lower half of the vector $(\phi,\psi)^t$ contains the values of the wave 
function on the boundaries, and the upper half contains the values of the 
normal derivative. 
Note that each boundary curve has two contributions to  
\equ~(\ref{eq:symbolic}) with identical $\psi$, $\phi$ (which are continuous
across the boundary) but different $B_j$, $C_j$.  

\subsection{Plane-wave scattering}
\label{sec:scattering}
The scattering states in the related quantum problem have been discussed again
by Knipp and Reinecke~\cite{KR96}. In contrast to the case of bounded states,
their results also apply to dielectric cavities.  

In region $\Omega_J$ the wave function has the asymptotic form as in
\equ~(\ref{eq:scatteringbc}). The incoming wave $\psi_{\text{in}}$ 
satisfies \equ~(\ref{eq:wave}). Thus, $\psi$ can 
be replaced by $\psi-\psi_{\text{in}}$ in
\equ~(\ref{eq:bim}) giving 
\begin{eqnarray}
\nonumber
\psi({\bf r}') & = &\exp{(i{\bf k}{\bf r}')} +
\oint_{\Gamma_J}ds\{[\psi(s)-\psi_{\text{in}}(s)]\partial_\nu G(s,{\bf r}';k)\\
\label{eq:scatteringbim}
& &- G(s,{\bf r}';k)[\phi(s)-\phi_{\text{in}}(s)]\} \ ,
\end{eqnarray}
where $\psi_{\text{in}}(s) = \exp{(i{\bf k}{\bf r})}$ and $\phi_{\text{in}}(s) = i{\bf k}{\bf
\nu}({\bf r})\exp{(i{\bf k}{\bf r})}$ at ${\bf r} = {\bf r}(s)$.
The circle at infinity does not contribute to the 
BIE~(\ref{eq:scatteringbim}) as in the case of bounded states. The reason,
however, is different as we shall see in greater detail in the following 
subsection when considering resonances.  

If ${\bf r}'$ is taken from the boundary then \equ~(\ref{eq:scatteringbim})
can be written as inhomogeneous integral equation
\begin{eqnarray}
\nonumber
\oint_{\Gamma_J}ds[B(s',s)\phi(s)+C(s',s)\psi(s)] & = & \\
\label{eq:hbimscatt}
\oint_{\Gamma_J}ds[B(s',s)\phi_{\text{in}}(s)+C(s',s)\psi_{\text{in}}(s)] & \ . &
\end{eqnarray}
Together with the other $J-1$ BIEs, which are of the same
form as in \equ~(\ref{eq:hbim}), the resulting inhomogeneous system 
of equations is 
\begin{equation}\label{eq:scatteringsymbolic}
M\left(\begin{array}{c}
  \phi\\
  \psi
 \end{array}
\right) =  
M_0\left(\begin{array}{c}
  \phi_{\text{in}}\\
  \psi_{\text{in}}
 \end{array}
\right)
\end{equation}
with 
\begin{equation}\label{eq:M_0}
M_0 = 
\left(\begin{array}{cc}
  0 & 0\\
  \vdots & \vdots\\
  0 & 0 \\
  B_J & C_J
 \end{array}
\right) \ .
\end{equation}

Having determined the solutions $\psi$ and $\phi$ we can compute the
differential scattering 
amplitude by evaluating \equ~(\ref{eq:scatteringbim}) for large $r'$ and 
comparing the result with \equ~(\ref{eq:scatteringbc})
giving 
\begin{eqnarray}
\label{eq:f}
f(\theta,{\bf k}) & = & \frac{1+i}{4\sqrt{\pi k}}
\oint_{\Gamma_J}ds \exp{[-i{\bf k}_f{\bf r}(s)]}\\
\nonumber
& & \{i{\bf k}_f{\bf \nu}(s)[\psi(s)-\psi_{\text{in}}(s)]+\phi(s)-\phi_{\text{in}}(s)\} \ ,
\end{eqnarray}
where ${\bf k}_f = (k\cos\theta,k\sin\theta)$ and $\theta$ is the detection
angle. Here, $|f(\theta,{\bf k})|^2$ is the differential scattering cross
section. The total cross section $\sigma({\bf k}) = \int d\theta |f(\theta,{\bf
k})|^2$ can be easily calculated from the 
forward-scattering amplitude, ${\bf k}_f = {\bf k} = (k\cos\phi,k\sin\phi)$, with the help of the
optical theorem (see, e.g., Ref.~\cite{Landau96})
\begin{equation}\label{eq:opticaltheorem}
\sigma({\bf k}) = 2\sqrt{\frac{\pi}{k}}\imagb{(1-i)f(\theta = \phi,{\bf k})} \
.  
\end{equation}

\subsection{Resonances}
\label{sec:resonances}
We now turn to the BIEs for resonances.
Comparing the scattering boundary condition~(\ref{eq:scatteringbc}) and the 
outgoing-wave condition~(\ref{eq:outgoingbc}) indicates that we possibly can 
use the scattering approach neglecting the incoming wave, that is 
\equ~(\ref{eq:scatteringsymbolic}) with $M_0 = 0$. Apart from the fact that
$k$ is now a complex number, this is then identical to 
\equ~(\ref{eq:symbolic}) for bounded states.
There is, however, one problem. The circle at infinity, $\Gamma_\infty$, may
give a nonvanishing contribution 
\begin{equation}\label{eq:Iinf}
I_\infty({\bf r}') = \oint_{\Gamma_\infty}ds[\psi(s)\partial_\nu
G(s,{\bf r}';k)- G(s,{\bf r}';k)\partial_\nu \psi(s)] 
\end{equation}
to the r.h.s. of \equ~(\ref{eq:bim}) because with $\imag{k} < 0$ neither the 
wave function~(\ref{eq:outgoingbc}) nor the Green's function~(\ref{eq:green2}) vanish at infinity.  
Gaspard and Rice~\cite{GR89} have shown for a Dirichlet scattering problem 
that nonetheless $I_\infty({\bf r}') = 0$ if ${\bf r}'$ is at one of the
scatterers' 
boundaries or if ${\bf r}'$ is at a large distance from these boundaries.   
We have to extend their result because (i) the problem of dielectric cavities 
involves a different kind of boundary conditions; (ii) we are interested in 
the wave function $\psi({\bf r}')$ also in the near-field.  
We start with recalling that the circle at infinity, $\Gamma_\infty$, is
defined by $r = \const$ with $r\to\infty$. Using the asymptotical
behaviour of Hankel functions of first 
kind~\cite{GradRyzh65} 
\begin{equation}\label{eq:hankel} 
H^{(1)}_m(z) \sim \sqrt{\frac{2}{\pi z}}\exp{[i(z-m\pi/2-\pi/4)]} 
\end{equation}
as $z=k|{\bf r}-{\bf r'}| \to\infty$, it can be shown that the Green's
function in 
\equ~(\ref{eq:green2}) is asymptotically given by
\begin{equation}\label{eq:asymgreen}
G({\bf r},{\bf r}';k) \sim g(\theta-\theta',r')\frac{\exp{(ikr)}}{\sqrt{r}} \ ,
\end{equation}
with 
\begin{equation}
g(\theta-\theta',r') = -\frac{1+i}{4\sqrt{\pi k}}\exp{[-ikr'\cos{(\theta-\theta')}]} \ .
\end{equation}
Equation~(\ref{eq:asymgreen}) has the same $r$-dependence as the 
outgoing-wave condition~(\ref{eq:outgoingbc}). With $G$ and $\psi$ 
appearing in \equ~(\ref{eq:Iinf}) in an antisymmetric way it follows
$I_\infty({\bf r}') = 0$ for all ${\bf r}'\in \Omega_J \cup \Gamma_J$. 
The fact that $I_\infty({\bf r}')$ vanishes for ${\bf r}'\in \Gamma_J$ means
that the BIEs~(\ref{eq:symbolic}) can indeed be used to determine the resonant 
wave numbers $k$. Moreover, since $I_\infty({\bf r}') = 0$ also for 
${\bf r}'\in \Omega_J$ \equ~(\ref{eq:bim}) can be used to compute the
corresponding wave functions in the entire domain. 
  
Having established that the resonances are solutions of the
BIEs~(\ref{eq:symbolic}) with complex-valued $k$, we now demonstrate that the  
BIEs~(\ref{eq:symbolic}) posses additional solutions which do not fulfil the 
outgoing-wave condition~(\ref{eq:outgoingbc}). We study this in an elementary
way for a single cavity of arbitrary shape. Outside this cavity sufficiently 
far away from its boundary, a solution of wave equation~(\ref{eq:wave}) can 
be expressed as 
\begin{equation}\label{eq:contwave}
\psi(r,\theta) = \sum_{m=-\infty}^\infty [\alpha_m^{(1)}H^{(1)}_m(kr)+\alpha_m^{(2)}H^{(2)}_m(kr)]\exp{(im\theta)} \ ,
\end{equation}
with Hankel functions of first and second kind~\cite{GradRyzh65} and with
unknown complex-valued parameters $\alpha_m^{(1)}$ and
$\alpha_m^{(2)}$. Without boundary conditions at infinity, solutions as in \equ~(\ref{eq:contwave})
exist for all values of $k$. Boundary conditions that fix all parameters 
$\alpha_m^{(2)}$ give rise to a discrete spectrum of $k$; for instance, the
outgoing-wave condition~(\ref{eq:outgoingbc}) requires $\alpha_m^{(2)} = 0$
for all $m$. Inserting the expansion~(\ref{eq:contwave}) into
\equ~(\ref{eq:Iinf}) leads to   
\begin{equation}\label{eq:intpro}
I_\infty({\bf r}') = 2\sum_{m=-\infty}^\infty \alpha_m^{(2)} J_m(kr')\exp{(im\theta')} \ .
\end{equation} 
Hence, $I_\infty({\bf r}')$ vanishes identically for all 
${\bf r}'\in \Omega_J \cup \Gamma_J$ only in the case of a resonance, where 
$\alpha_m^{(2)} = 0$ for all $m$.

However, the circle at infinity does not contribute to the 
BIEs~(\ref{eq:symbolic}) already if the weaker condition $I_\infty({\bf r}') =
0$ for ${\bf r}'\in \Gamma_J$ is satisfied. We insert this condition into the
l.h.s. of \equ~(\ref{eq:intpro}) and note that the r.h.s. is an expansion of
a solution of wave equation~(\ref{eq:wave}) inside the cavity with 
``wrong'' index of refraction $n = n_J = 1$. 
The result is that the BIEs~(\ref{eq:symbolic}) possess undesired solutions,
namely bounded states of an interior Dirichlet problem, in addition to the
resonances. As one consequence, the solutions of the scattering
BIEs~(\ref{eq:scatteringsymbolic}) are not unique whenever $k$ is a solution of
the interior Dirichlet problem. Note that this nonuniqueness has not been
discussed by Knipp and Reinecke~\cite{KR96}.

A related problem is known for cases with Dirichlet or Neumann
conditions; see, e.g., Refs.~\cite{CB91,CZ92}.  
There have been several attempts to modify the BIEs in order to get
rid of these ``spurious solutions''. Some of these modifications could be
applied to the present case, but this would result in singular integrals which
are hard to deal with numerically.
Fortunately, the spurious solutions are not a severe problem for our
purpose. We can distinguish them, in principle, from the resonances in which
we are interested in. The former have $\imag{k} = 0$ whereas the latter have
$\imag{k} < 0$.

\subsection{TE polarization}
\label{sec:TE}
In the case of TE polarization, \equ~(\ref{eq:wave}) is valid
with $\psi$ representing the magnetic field $H_z$. The wave function $\psi$ is
continuous across the boundaries, but its normal derivative is not, in
contrast to the case of TM polarization. Instead, $n({\bf
r})^{-2}\partial_\nu\psi$ is continuous~\cite{Jackson83}. 

This new boundary condition can be easily incorporated in the BEM by defining
$\phi = n^{-2}\partial_\nu\psi$, $B(s',s) = -2G(s,s';k)n^2$ and
$\phi_{\text{in}}$ accordingly in equations like \eqs~(\ref{eq:hbim}) and
(\ref{eq:hbimscatt}). We remark that the spurious solutions are not affected by
this change of boundary conditions. 

\subsection{Symmetry considerations}
\label{sec:symmetry}
\begin{figure}[ht]
\includegraphics[width=4.5cm,angle=0]{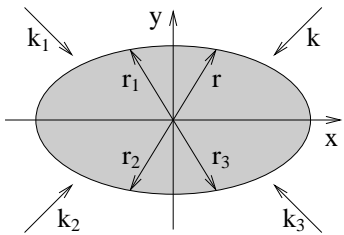}
\caption[]{\footnotesize Symmetric cavity.}
\label{fig:symmetry}
\end{figure}

Many dielectric cavities studied in the literature possess discrete
symmetries. For example, the elliptical cavity in \fig~\ref{fig:symmetry} is
symmetric with respect to the $x$ and $y$ axes. In such a case, the wave 
functions can be divided into four symmetry classes 
\begin{eqnarray}
\label{eq:symmetryconditionsa}
\psi_{\zeta\xi}(-x,y) & = & \zeta\psi_{\zeta\xi}(x,y) \ ,\\ 
\label{eq:symmetryconditionsb}
\psi_{\zeta\xi}(x,-y) & = & \xi\psi_{\zeta\xi}(x,y) \ ,
\end{eqnarray}
with the parities $\zeta \in \{-,+\}$ and $\xi \in \{-,+\}$. The normal
derivative obeys the same symmetry relations.

For systems with symmetries the BIEs can be reduced to a fundamental domain if
a modified Green's function is used. This decreases the numerical effort
considerably. Let us restrict our discussion to the case in 
\eqs~(\ref{eq:symmetryconditionsa}) and (\ref{eq:symmetryconditionsb}); other 
symmetries can be treated in a similar way.
The BIEs~(\ref{eq:cauchybim}) reduce to integrals along the boundaries
restricted to the quadrant $x,y \geq 0$ if the Green's function $G({\bf
r},{\bf r}')$ is replaced by
\begin{equation}
G({\bf r},{\bf r}')+\zeta G({\bf r}_1,{\bf r}')
+\zeta\xi G({\bf r}_2,{\bf r}')+\xi G({\bf r}_3,{\bf r}')
\end{equation}
with ${\bf r} = (x,y)$, ${\bf r}_1 = (-x,y)$, ${\bf r}_2 = (-x,-y)$, 
${\bf r}_3 = (x,-y)$; see \fig~\ref{fig:symmetry}. 
The derivative $\partial_\nu G(s,{\bf r}')$ is modified in the same 
way with the normal unit vector ${\bf \nu}$ changing as ${\bf r}$. 

The scattering problem as formulated in \sect~\ref{sec:scattering} does not
allow the symmetry reduction because the incoming plane wave in general
destroys the symmetry; $\phi_{\text{in}}$ and $\psi_{\text{in}}$
in \equ~(\ref{eq:scatteringsymbolic}) do not fulfil the 
conditions~(\ref{eq:symmetryconditionsa}) and (\ref{eq:symmetryconditionsb}).
There are certain incoming directions which do not spoil the
symmetry, but using only these special directions is dangerous because
possibly not all resonances are excited. 
A better approach is to consider a different physical situation illustrated in
\fig~\ref{fig:symmetry}. Four plane waves are superimposed to a symmetric
incoming wave 
\begin{equation}
\psi_{\text{in}} = \exp{(i{\bf k}{\bf r})} + \zeta \exp{(i{\bf k}_1{\bf r})} 
+ \zeta\xi\exp{(i{\bf k}_2{\bf r})} + \xi \exp{(i{\bf k}_3{\bf r})}
\end{equation}
where ${\bf k} = (k_x,k_y)$, ${\bf k}_1 = (-k_x,k_y)$, ${\bf k}_2 =
 (-k_x,-k_y)$, ${\bf k}_3 = (k_x,-k_y)$. 
With this incoming wave, the scattering problem can be symmetry reduced.  
A more general formulation for an arbitrary symmetry can be found in
Ref.~\cite{ND94}.   

\section{Boundary element method}
\label{sec:bem}
The most convenient numerical strategy for solving BIEs as in
\eqs~(\ref{eq:hbim}) and (\ref{eq:hbimscatt}) is the BEM. The boundary
is discretized by dividing it into small boundary elements. Along such an
element, the wave function and its normal derivative are considered as being
constant (for linear, quadratic, and cubic variations see, e.g., 
Refs.~\cite{CB91,Banerjee94}). Equation~(\ref{eq:hbim}) is therefore 
approximated by a sum of $N_j$ terms
\begin{equation}
\sum_{l=1}^{N_j} (B_{il}\phi_l+C_{il}\psi_l) = 0
\end{equation}
where $B_{il} = \int_l ds\, B(s_i,s)$, $C_{il} = \int_l ds\, C(s_i,s)$,
$\phi_l = \phi(s_l)$, $\psi_l = \psi(s_l)$, and $\int_l$ denotes the
integration over a boundary element with midpoint $s_l$.  
The entire set of BIEs is approximated by an equation as in 
\equ~(\ref{eq:symbolic}), but for which $B_j$ and $C_j$ are $N_j\times N$
matrices, $M$ is a $2N\times 2N$ (non-Hermitian complex) matrix, $\phi$ and 
$\psi$ are $N$-component vectors with $2N = \sum_{j=1}^J N_j$. 
Note that each boundary element belongs to two different regions.
In the same way the scattering problem is approximated by an equation as in 
\equ~(\ref{eq:scatteringsymbolic}) with $M_0$ being a $2N\times 2N$ 
matrix, $\phi_{\text{in}}$ and $\psi_{\text{in}}$ being $N$-component
vectors.   

In the literature several levels of approximation are used for the matrix
elements $B_{il}$ and $C_{il}$. The crudest approximation is to
evaluate such an integral only at the corresponding midpoint $s_l$. While this
is sufficient for the calculation of bounded states in quantum 
billiards~\cite{Baecker02}, in our case the small imaginary parts of $k$
require a more accurate treatment. We
therefore do perform the numerical integration of the
matrix elements $B_{il}$ and $C_{il}$, using standard integration routines
like, for example, Gaussian quadratures~\cite{Press88}. The number of interior
points in the range of integration should be chosen large if the boundary
elements $s_i$ and $s_l$ are close to each other and small if they are far
away.  
Moreover, our experience is that the results are considerably more accurate 
if the boundary elements are not approximated by straight lines but, instead, 
the exact shape of the boundary elements is used for all interior
points in the range of integration. 

Due to the almost singular behaviour of the integral kernels at ${\bf r}'
= {\bf r}$, the diagonal elements $C_{ll}$ and $B_{ll}$ require special
care. Inserting the limiting cases for small boundary-element length $\Delta
s_l$ in \eqs~(\ref{eq:H11small}) and (\ref{eq:cosalphakappa}) into
\equ~(\ref{eq:normalgreen}) gives 
\begin{equation}
C_{ll} =  -1 + \frac{\kappa_l}{2\pi}\Delta s_l\ ,
\end{equation}
where $\kappa_l$ is the curvature at point $s_l$.
To approximate $B_{ll}$ accurately, more higher order terms than in
\equ~(\ref{eq:H10small}) are needed:
\begin{equation}
H^{(1)}_0(z) \sim \frac{2i}{\pi}\ln{\frac{z}{2}} +1+\frac{2i}{\pi}\gamma
\ ,
\end{equation}
where $\gamma = 0.577215\ldots$ is Euler's constant.
Integration yields
\begin{equation}
B_{ll} = \frac{\Delta s_l}{\pi}[1-\ln{\frac{n_jk\Delta s_l}{4}}+i\frac{\pi}{2}-\gamma]
\ .
\end{equation}

\subsection{Treatment of corners}
\label{sec:corners}
\begin{figure}[ht]
\includegraphics[width=5.5cm,angle=0]{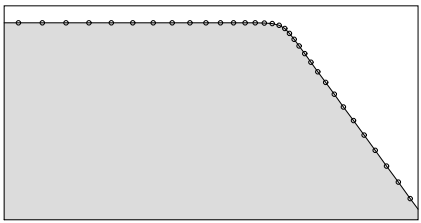}
\caption[]{\footnotesize Rounded corner. The number of discretization points
(circles) is enhanced at the corner.}
\label{fig:corner}
\end{figure}

Dielectric corners are numerically difficult to treat because certain
components of the electric field can be infinite at the corner (see the
discussion in Ref.~\cite{Hadley02} in the context of dielectric waveguides).
In the BEM, a corner leads to a second problem.
The integral kernel of $C_{ll}$ has a singularity caused by a
diverging curvature $\kappa$; see \equ~(\ref{eq:cosalphakappa}). 
To circumvent these difficulties, we
smooth the boundary as sketched in \fig~\ref{fig:corner}. The curvature
$\kappa$ and the electric field are then everywhere bounded.  

The minimum value of the radius of curvature, $\rho = 1/\kappa$, along such a 
rounded corner should be much larger than the typical distance 
between discretization points, so that the boundary is locally
smooth. However, in order to ensure that the rounding does not influence the 
result, $\rho$ should be much smaller than the wavelength $\lambda$. 
Clearly, these requirements can be met most efficiently by using a nonuniform 
discretization with a relatively large density of discretization points at
corners as illustrated in \fig~\ref{fig:corner}. Since the results do
not depend on the particular selected rounding and discretization we do not
give explicit formulae.

\subsection{Finding and computing resonances}
The scattering problem as discussed in \sect~\ref{sec:scattering} provides us
with first approximations to the wave numbers of the resonances. 
Let us fix $\phi$ to an appropriate value and plot the total cross section in
\equ~(\ref{eq:opticaltheorem}) as function of $k$ in the range of interest. 
Resonances can be identified as peaks. The peak position $\alpha$ and the
width $\gamma$ determine the resonant wave number as $k_{\text{res}} \approx
k_1 = \alpha-i\gamma/2$.  
It might be difficult to resolve numerically very broad and very narrow peaks, 
because they are hidden either in the background or between two consecutive 
grid points. For microlaser operation, however, these two extreme cases are
not relevant. Too short-lived resonances (broad peaks) fail to provide a
sufficient long lifetime  for the light to accumulate the gain required to
overcome the lasing threshold, whereas too long-lived resonances (narrow
peaks) do not supply enough output power.  

The spurious solutions of the interior Dirichlet problem
occasionally appear in the scattering spectrum as extremely narrow peaks. The
reason is that numerical inaccuracies broaden the $\delta$-peaks to peaks of
finite width. However, choosing a sufficiently fine boundary discretization 
and/or an appropriate, not too fine discretization in $k$ reduces the
probability of observing them. Moreover, they can be removed with a simple
trick: use $k$ with a small negative imaginary part in
\equ~(\ref{eq:opticaltheorem}).    

The discretized version of \equ~(\ref{eq:symbolic}) has a nontrivial solution 
only if $\det{M(k_{\text{res}})} = 0$. Using a first approximation $k_1$ from
the scattering problem as starting value, we find a much better approximation 
to $k_{\text{res}}$ in the complex plane with the help of Newton's
method  
\begin{equation}\label{eq:newton}
k_{l+1} = k_l-\frac{g(k_l)}{g'(k_l)} 
\end{equation}
with $l = 1,2,\ldots$ and $g(k)=\det{M(k)}$. The derivative $g'(k) = \partial
g(k)/\partial k$ can be approximated by  
\begin{equation}\label{eq:diffquo}
g'(k) \approx  \frac{g(k+\Delta)-g(k)}{2\Delta}
-i \frac{g(k+i\Delta)-g(k)}{2\Delta} \ ,
\end{equation}
where $\Delta$ is a small real number. Equation~(\ref{eq:newton}) is repeated
iteratively until a chosen accuracy is achieved.

Newton's method in \equ~(\ref{eq:newton}) is very efficient close to an
isolated resonance where $\det{M} \propto k-k_{\text{res}}$. For 
$q$-fold degenerate resonances the determinant behaves like
$(k-k_{\text{res}})^q$. The resulting problem of slow convergence can
be eliminated by choosing $g=(\det{M})^{1/q}$.   

A slightly different approach for finding resonances can be gained by
 rewriting Newton's method in \equ~(\ref{eq:newton}) with the help of the 
matrix identity $\ln\det{M} = \text{tr}\ln{M}$ as
\begin{equation}\label{eq:newtontrace}
k_{l+1} = k_l-\frac{q}{\text{tr}[M^{-1}(k_l)M'(k_l)]} \ ,
\end{equation}
where $\text{tr}$ denotes the trace of a matrix.
The derivative $M'(k)$ can be calculated as in \equ~(\ref{eq:diffquo}). It
turns out that the numerical algorithm corresponding to 
\equ~(\ref{eq:newtontrace}) is a bit faster than the original Newton's
method in \equ~(\ref{eq:newton}). 

Having found a particular wave number $k_{\text{res}}$, the vector components
$\phi_l$ and $\psi_l$ are given by the null eigenvector of the square 
matrix $M(k_{\text{res}})$. This eigenvector can be found with, for instance, 
singular value decomposition~\cite{Press88}. 
The wave function in each domain $\Omega_j$ is then constructed by
discretizing \equ~(\ref{eq:bim})
\begin{eqnarray}\label{eq:bimdiscretized}
\psi({\bf r}') & = & \sum_l \psi_l \int_l ds\,\partial_\nu
G(s,{\bf r}';k_{\text{res}}) \\
\nonumber
& & - \sum_l \phi_l \int_l ds\, G(s,{\bf
r}';k_{\text{res}}) \ ,
\end{eqnarray}
where $l$ runs over all boundary elements of $\Gamma_j$.

How fine must be the discretization of the boundary in order to obtain a good
approximation of a resonance at $k_{\text{res}}$? The local wavelength $\lambda =
2\pi/n\real{k_{\text{res}}}$ is the smallest 
scale on which the wave function and its derivative may vary. Hence, the 
minimum number of boundary elements along each wavelength, $b =
\lambda/\Delta s$, should be larger or equal than at least $4$; $\Delta s$ is
the maximum value of all lengths $\Delta s_i$.
We have verified the BEM using different values of $b$. Taking $b =
16$, we find good agreement with the separation-of-variables solution of the
circular cavity (see e.g. Ref.~\cite{BarberHill90}) and to results of the
wave-matching method obtained for the quadrupolar cavity~\cite{ND97}. 
Only for extremely long-lived resonances larger $b$ are necessary to 
determine the very small imaginary parts of $k$ accurately (recall that this 
is important for distinguishing spurious solutions from real resonances). 
However, as already explained, extremely long-lived resonances are not
relevant for microlaser applications and, moreover, they occur only in
circular or slightly deformed circular cavities for which the wave-matching
method is more suitable anyway.   

\section{Example: two coupled hexagonal-shaped cavities}
\label{sec:example}
Vietze {\it et al.} have experimentally realized hexagonal-shaped microlasers 
by putting laser active dyes into molecular sieves made of 
${\text{AlPO}_4-5}$~\cite{VKLISLA98}. 
Numerical simulations on rounded hexagons based on the wave-matching method
have shown convergence problems at corners~\cite{BILNSSVWW00,Noeckelpc02}.  
The following example is relevant for future experiments and demonstrates
that the BEM can handle arbitrarily 
sharp corners and, moreover, coupled resonators. 
Near-field-coupling of resonators is interesting, because it may improve the
optical properties of the resonators, as e.g. the far-field directionality.

\begin{figure}[ht]
\includegraphics[width=6.0cm,angle=0]{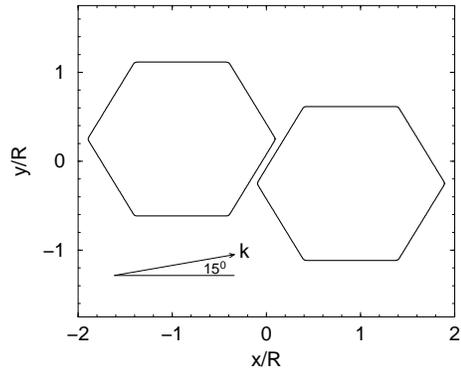}
\caption[]{\footnotesize Two hexagonal cavities. The incoming plane wave with
wave vector ${\bf k}$ is incidence at $15^\circ$ to the horizontal side faces.}
\label{fig:example}
\end{figure}

Figure~\ref{fig:example} illustrates the configuration: two hexagonal cavities
with sidelength $R$ are displaced by the vector $(1.8R,0.5R)$. According to the
experiments in Ref.~\cite{VKLISLA98,BILNSSVWW00}, the polarization is of TM
type, the index of refraction is
$n=1.466$ inside the cavities and $n=1$ outside; $R$ ranges from $4\mu\text{m}$
to $10\mu\text{m}$, the wavelength $\lambda$ from $600\text{nm}$ to
$800\text{nm}$ depending on the dye. Since only the ratio between $R$ and 
$\lambda$ is relevant, we use in the following the dimensionless wave
number $kR$. 
We focus on a $kR$-interval from $20$ to $25$ within the experimental 
spectral interval. A total of $2N=3200$ 
discretization points is then sufficient. We slightly smooth the corners as
discussed in \sect~\ref{sec:corners} such that $\rho/\lambda \approx 0.11$ and
$\rho/\Delta s\approx 11.2$. 

\begin{figure}[ht]
\includegraphics[width=6.0cm,angle=0]{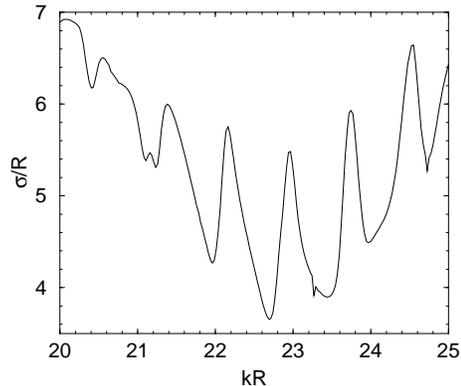}
\caption[]{\footnotesize Calculated total cross section $\sigma/R$ vs. $kR$ for
two coupled hexagonal resonators. The plane wave is incidence at $15^\circ$ to
the horizontal side faces; cf. \fig~\ref{fig:example}.} 
\label{fig:sigma}
\end{figure}

Figure~\ref{fig:sigma} shows the total cross section $\sigma$ for plane-wave
scattering with incidence angle $\phi = 15^\circ$ computed from
\equ~(\ref{eq:opticaltheorem}). The dominant structure is a series of
equidistant peaks of roughly Lorentzian shape. At $kR \approx 23.25$ 
we identify a spurious solution of the
interior Dirichlet problem. The fact that it is the only one visible in the
chosen range of wave numbers confirms that the spurious solutions 
are not a problem.

\begin{figure}[ht]
\includegraphics[width=7.5cm,angle=0]{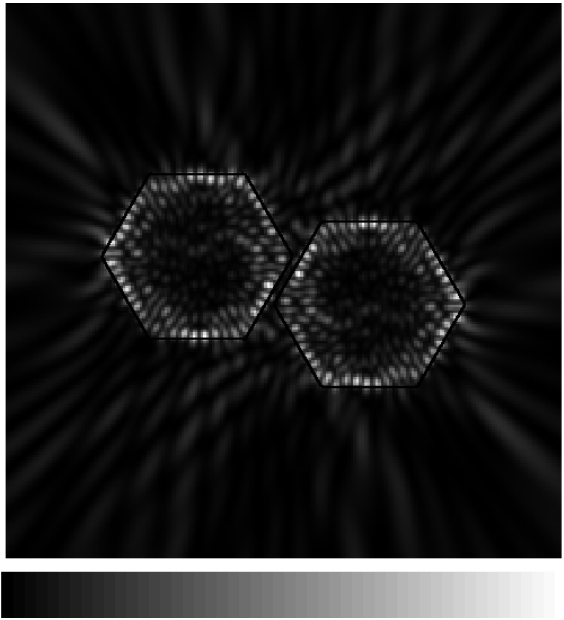}
\caption[]{\footnotesize Calculated near-field intensity pattern $|\psi({\bf r})|^2$ of 
the resonance with $k_{\text{res}}R \approx 22.94444-i0.09696$. Intensity is higher for light
regions and vanishes in the black regions.} 
\label{fig:resonance}
\end{figure}

\begin{figure}[ht]
\includegraphics[width=7.0cm,angle=0]{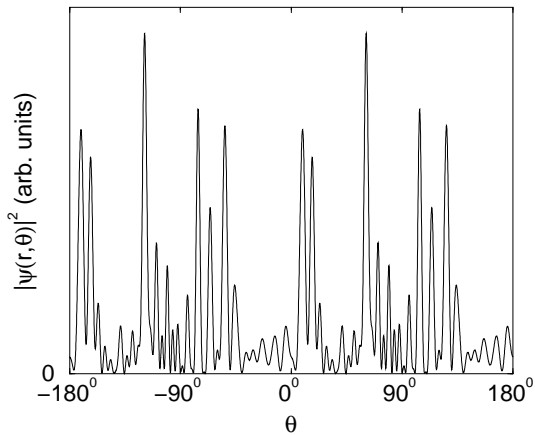}
\caption[]{\footnotesize Far-field emission pattern, $|\psi(r,\theta)|^2$ with
large $r$, of the resonance in \fig~\ref{fig:resonance}.} 
\label{fig:farfield}
\end{figure}

The peak at $kR \approx 22.95$ in \fig~\ref{fig:sigma} has roughly the width
$0.196$, so 
we use $k_1R = 22.95-i0.098$ as initial guess for Newton's method in \equ~(\ref{eq:newtontrace}). The more precise location of 
the resonance is found to be $k_{\text{res}}R \approx 22.94444-i0.09696$. The near-field intensity
pattern in \fig~\ref{fig:resonance} and the far-field emission pattern in 
\fig~\ref{fig:farfield} are computed with the help of \equ~(\ref{eq:bimdiscretized}). 
A detailed account of the structure of this kind of resonances and its
implication on the properties of the microlasers will be given in a
future publication.    

\section{Summary}
\label{sec:summary}
We have introduced a boundary element method (BEM) to compute TM and TE
polarized resonances with intermediate lifetimes in dielectric cavities. 
We have discussed spurious solutions, the treatment of
cavities with symmetries and cavities with corners.
Numerical results are shown for an example of two coupled
hexagonal cavities.  

If compared to finite-difference methods and related methods the BEM is very
efficient since the wave function and its derivative are only evaluated at the
boundaries of the cavities.    
It is in general less efficient than the wave-matching method but in contrast
to the latter it can be applied to complex geometries, such as cavities with
corners and coupled cavities. 

The BEM is especially suitable for computing phase space representations 
of wave functions such as the Husimi function which also only requires the wave
function and its normal derivative on the domain boundaries~\cite{HSS02}. 

\begin{acknowledgments}
I would like to thank M. Hentschel, S. W. Kim, J. N\"ockel, F. Laeri and A. B\"acker for discussions.
The work was supported by the Volkswagen foundation (project
``Molekularsieblaser-Konglomerate im Infraroten'').
\end{acknowledgments}

\bibliographystyle{prsty}
\bibliography{}

\end{document}